\documentclass[showpacs,amsmath,superscriptaddress,pre]{revtex4}
\usepackage{graphicx}
\usepackage{amsmath}
\usepackage{epsfig}
\usepackage{graphics}
\usepackage{color}
\usepackage{latexsym}
\usepackage{amsfonts}
\usepackage{amssymb}
   
\usepackage{plain}

\begin{document}

\title{Spontaneous imbibition in a slit pore: a lattice-gas  dynamic mean field study}
\author{E. Kierlik}
\affiliation{Laboratoire de Physique Th\'eorique de la Mati\`ere Condens\'ee, Universit\'e Pierre et Marie Curie\\ 4 place Jussieu, 75252 Paris Cedex 05, France}
\author{F. Leoni}
\affiliation{GIT-SPEC, CEA Saclay, 91191 Gif-sur-Yvette Cedex, France}
\author{M. L. Rosinberg}
\affiliation{Laboratoire de Physique Th\'eorique de la Mati\`ere Condens\'ee, Universit\'e Pierre et Marie Curie\\ 4 place Jussieu, 75252 Paris Cedex 05, France}
\author{G. Tarjus}
\affiliation{Laboratoire de Physique Th\'eorique de la Mati\`ere Condens\'ee, Universit\'e Pierre et Marie Curie\\ 4 place Jussieu, 75252 Paris Cedex 05, France}



\begin{abstract}
We present a theoretical study of spontaneous imbibition in a slit pore using a lattice-gas model and a dynamic mean-field theory. Emphasis is put on the influence of the precursor films on the speed of the imbibition front due to liquid mass conservation. This work is dedicated to Bob Evans for his 65th birthday in recognition of his seminal contributions to the theory of fluids in confining geometries. 
\end{abstract}

\maketitle

\def\be{\begin{equation}}
\def\ee{\end{equation}}
\def\bea{\begin{align}}
\def\eea{\end{align}}

\section{Introduction}

Spontaneous imbibition, namely the rise of a liquid in a capillary tube and more generally in a porous solid, is a ubiquitous phenomenon that has received a lot of attention over the years because of its crucial importance in many industrial processes (oil recovery, ink printing, textile dyeing, etc...) as well as in agriculture and biological sciences.  In spite of the complexity of real porous media (in particular their disordered structure\cite{ADR2004}), it is often observed that the distance of penetration of the liquid inside the solid increases asymptotically as the square root of time, as predicted by the classical Lucas-Washburn (LW) equation\cite{L1918} that describes  the fluid behavior in a single capillary.  On the macroscopic scale, the $\sqrt{t}$ law results from the balance between a constant capillary driving force due to the pressure drop across the liquid/vapor meniscus and an  increasing viscous drag.  In recent years, in relation to the rapid development of nanofluidic devices, the applicability of these macroscopic concepts in strongly confined geometries has attracted much interest, on both the experimental\cite{HGSKV2007} and theoretical\cite{DMB2007, C2008, CBD2008, KPGPY2008, AK2009,WSP2010,SLCB2010} sides. The main conclusion of these studies is that the LW equation still works in capillaries with diameters of a few molecular sizes, the prediction being almost quantitative in the case of a simple Lennard-Jones fluid\cite{DMB2007}. This is quite remarkable as the continuum hydrodynamic description of the fluid is expected to break down at very small length scales. 

In the case of complete wetting, molecular dynamics (MD) and lattice-Boltzmann simulations\cite{DMB2007,C2008,CBD2008} have also reported the presence of microscopic films moving ahead of the main capillary front and following also a $\sqrt{t}$ law, but with a different prefactor than that of the main front. Such precursor films are commonly observed in spreading droplet experiments  and several theoretical models have been proposed to reproduce their diffusive behavior\cite{BEIMR2009}. In the case of imbibition, wetting films can significantly affect the dynamics of the gas-liquid interface\cite{BQ2003} and it is generally expected that they reduce the viscous drag, which of course is a crucial issue in microfluidics. Some aspects of the problem, however, are not as well understood and deserve more systematic investigations. In particular, one would like to better understand the experimental conditions required for the appearance of the films, the influence of the solid wettability on their behavior, and how they compete with the meniscus for the liquid coming from the reservoir. 

These are complex issues, due to the coupling of hydrodynamic and non-hydrodynamic modes in the vicinity of the contact line.  Ideally, a theoretical study should require a full atomistic description of the system, but the very different length scales involved in the problem make it  computationally very demanding. An important simplification consists in adopting a coarse-grained (or mesoscopic) lattice-gas description, as is done in the lattice-Boltzmann method which indeed appears as an efficient tool for modelling capillary filling\cite{C2008,CBD2008,KPGPY2008}. In the present work, we further simplify the problem by treating the dynamical evolution of the fluid in configurational space only, the effect of the hydrodynamic modes being simply incorporated in effective parameters: only diffusionlike mechanisms are explicitly included. This minimal model, however, does account for liquid conservation which is the key ingredient at the origin of the constant slowing down of the imbibition front.  A similar approach underlies the recent phase-field models of imbibition in disordered porous solids\cite{ADR2004,DREAMA2000,PH2006}. The study of such systems is the next step on our agenda\cite{LKRT2010}, and its theoretical treatment indeed requires the above simplifications.  Here,  we describe spontaneous imbibition in a slit-pore by using a dynamic (lattice) mean-field theory (DMFT) which is essentially the mean-field version of the Kawasaki spin exchange dynamics. This theory is consistent with the treatment of thermodynamics at the mean-field level and it has already proven useful to model the dynamics of adsorption/desorption phenomena in nanopores\cite{M2008,EM2009,EM2010}.  

Although we shall mainly consider the case of primary imbibition, when a liquid invades a dry capillary and precursor films appear in the complete wetting regime (which is the situation considered in the recent simulation studies\cite{DMB2007,CBD2008}), we shall also briefly investigate the case where the liquid advances over pre-existing thin films (prewetted capillary) which is encountered in many actual situations, in particular in experiments with nanoporous solids\cite{HGSKV2007,N2006}. 

\section{Model and theory}

\begin{figure}[hbt]
\begin{center}
\includegraphics[width=9cm]{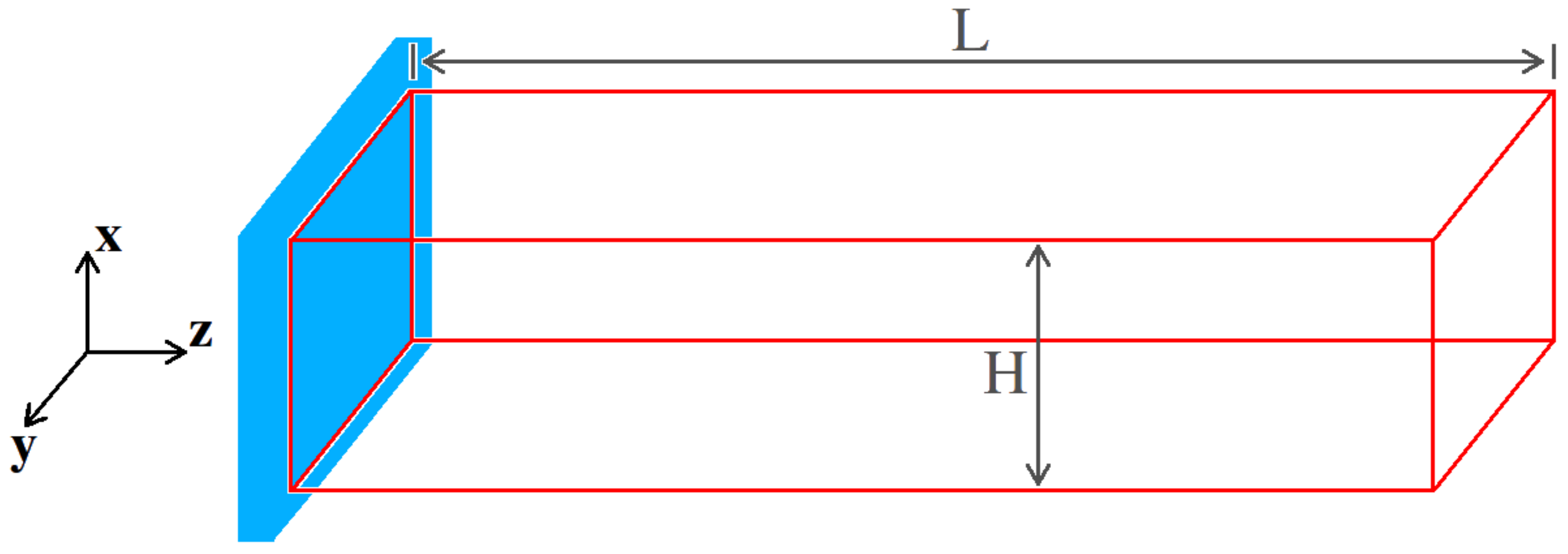}
\caption{\label{Fig1} (Color on line) Schematic representation of the slit pore geometry used in the DMFT calculations. The pore has  a length $L$ is the $z$ direction and a width $H$ in the $x$ direction. Periodic boundary conditions are applied in the $y$ direction. The liquid reservoir (in blue) is located at the left side of the pore.}
\end{center}
\end{figure}

We consider the case of a slit pore schematically represented in Fig. \ref{Fig1}. 
Before describing the dynamic mean-field approach, we first briefly recall the macroscopic description that leads to the LW $\sqrt t$ law.  As pointed out in the introduction, the dynamics of spontaneous imbibition under steady conditions (and neglecting evaporation, gravity, and inertial effects\cite{ADR2004}) is governed by the balance between the capillary driving force due to the Laplace pressure across the liquid/gas meniscus and the viscous drag of the liquid.  In a slit geometry the meniscus is cylindrical and the Laplace pressure is given by
\begin{equation}
\label{EqLap}
P_c=\frac{2\gamma_{lg}\cos \theta}{H}
\end{equation}
where $H$ is the width of the slit, $\gamma_{lg}$ is the surface tension of the liquid-gas interface and $\theta$ is the contact angle between the liquid and the solid walls (assuming that in the long-time limit one can replace the dynamic contact angle by its static value). For an incompressible fluid, this gives rise to a constant gradient pressure over the whole imbibed region,
\begin{equation}
\Delta P=-\frac{P_c}{h(t)}
\end{equation}
where $h(t)$ is the position of the meniscus inside the pore at time $t$ (neglecting the internal structure of the interface). 
For a parabolic fluid-velocity profile normal to the walls (Hagen-Poiseuille laminar flow), the viscous drag implies that the average velocity of the front is proportional to $\Delta P$, 
\begin{equation}
\dot h(t)=-\frac{\kappa}{\eta }\Delta P
\end{equation}
where $\eta$ is the fluid viscosity and $\kappa=H^2/12$ is the pore permeability. This leads to the differential equation
\begin{equation}
h\dot h=\frac{\gamma_{lg}H\cos \theta }{6\eta}\, ,
\end{equation}
which after integration gives
\begin{equation}
\label{EqLW}
h(t)-h(t=0)=\big(\frac{\gamma_{lg}H\cos \theta}{3\eta}\big)^{1/2} t^{1/2} .
\end{equation}
The progression of the liquid inside the pore is thus described by  a diffusive law with an effective  diffusion coefficient that scales like the pore width.  Since the viscous drag is smaller in larger pores,  the liquid progresses faster, as indeed observed experimentally.

Our presentation of DMFT closely follows that of Refs.\cite{M2008,EM2009,EM2010} and we refer the reader to these recent papers for more details (see also Ref. \cite{MAD2004} and the comprehensive review\cite{GPDM2003}). This mean-field version of the Kawasaki spin exchange model was first applied to the study of phase separation and surface enrichment in solid binary alloys\cite{B1974}. More generally, it is appropriate to describe systems with a diffusionlike behavior. The fluid inside the pore is modeled as a simple cubic lattice gas with Hamiltonian
\begin{equation}
\label{Eq1}
\mathcal{H}=-w_{ff}\sum_{<ij>}n_{i} n_{j}-w_{sf}\sum_{i,\:  surface}n_{{\bf i}}
\end{equation}
where $n_i$  is the occupancy variable at site $i$ ($n_i=0,1$) and the first sum runs over all nearest-neighbor pairs associated with fluid sites while the second sum runs over the fluid sites within the layers adjacent to the two walls (the pore walls being aligned with the $(100)$ planes of the cubic lattice); $w_{ff}$ and $w_{sf}$ denote the nearest-neighbor fluid-fluid and solid-fluid attractive interactions, respectively, and the wettability of the pore walls is thus controlled by the parameter $\alpha=w_{sf}/w_{ff}$. The corresponding bulk lattice gas has a symmetric coexistence curve $\rho_l(T) = 1 - \rho_g(T)$ with chemical potential $\mu_{sat}=-3w_{ff}$, independent of temperature T.

This type of lattice-gas model has been extensively used in the literature to study adsorption phenomena and phase transitions on substrates and in confined geometry (see {\it e.g.} \cite{OG1978,PSW1982,NF1983,BME1987}). Here, for simplicity, the effect of the solid is felt only in the layers immediately adjacent to the pore walls, as in the recent applications of DMFT to capillary condensation and evaporation\cite{M2008,EM2009,EM2010}. This is of course a simplification of the actual physical situation, where one typically has a long-ranged van der Waals interaction which decays as a power-law with the distance from the surface\cite{EM1984}.  Note also that we do not include any effect of surface corrugation in our calculation, although they may play an important role at low temperature. These two effects can be introduced in the theoretical treatment at the price of increasing numerical complexity. In the present set-up, the system is homogeneous in the $y$ direction parallel to the walls and is effectively two-dimensional.

Within the master equation approach\cite{B1974}, one can write down an exact equation for the temporal evolution of the mean site occupancy variable\cite{GPDM2003,M2008,EM2009,EM2010,MAD2004} 
\begin{equation}
\rho_i(t)=\langle n_i\rangle_t=\sum_{\{n\}}n_iP(\{n\},t),
\end{equation}
where $P(\{n\},t)$ is the probability to find the occupancy configuration $\{n\}$ at time $t$. Assuming that the transport is due to a hopping process between nearest-neighbor sites corresponding to a Kawasaki exchange dynamics, this evolution equation can be written as
\begin{equation}
\label{continuity}
\frac{\partial\rho_i}{\partial t}=-\sum_{j/i}J_{ij}(t)
\end{equation}
where the summation runs over nearest neighbors of site $i$ and the flux $J_{ij}(t)$ from site $i$ to site $j$ is given by 
\begin{equation}
J_{ij}(t)=\langle J_{ij}(\{n\})\rangle_t=\langle
w_{ij}(\{n\})n_i(1-n_j)-w_{ji}(\{n\})n_j(1-n_i)\rangle_t \, ,
\end{equation}
with $w_{ij}(\{n\})$ the transition probability for transitions from site $i$ to site $j$ for a configuration $\{n\}$. Through the Monte Carlo method with Metropolis transition probabilities, one could obtain an explicit realization of this dynamical process. In the DMFT, one instead uses a mean-field approximation and merely replaces the occupancy variables by their ensemble average (which  amounts in particular to assuming that the occupancy variables are dynamically uncorrelated). This yields
\begin{equation}
J_{ij}(t)=w_{ij}\rho_i(1-\rho_j)-w_{ji}\rho_j(1-\rho_i) \ ,
\end{equation}
and the evolution equation for the site densities becomes
\begin{equation}
\label{evol}
\frac{\partial\rho_i}{\partial t}=-\sum_{j/i}\left[w_{ij}\rho_i(1-\rho_j)-w_{ji}\rho_j(1-\rho_i)\right] 
\end{equation}
with transition probabilities 
\begin{equation}
\label{wij}
w_{ij}(\{\rho\})=w_0\exp(-\beta E_{ij})
\end{equation}
where
\begin{equation}
E_{ij}=\left\{
\begin{array}{ll}
0,       & E_j<E_i\\
E_j-E_i, & E_j>E_i
\end{array}\right.
\end{equation}
and
\begin{equation}
E_i=-w_{ff}\sum_{j/i}\rho_j-\left\{
\begin{array}{ll}
0     & \mbox{if i is a ``bulk" site}\\
w_{sf}  &\mbox{if i is a ``surface" site}.
\end{array}\right. 
\end{equation}
The parameter $w_0$ is an elementary jump rate which sets the time scale. We take the same $w_0$  for the ``bulk" sites and the ``surface" sites adjacent to the pore walls, thereby implying that the diffusion coefficient for a single molecule does not depend on the solid-fluid interaction, which is an additional approximation\cite{VVOCC1998}. The above dynamics could also be interpreted as resulting from a coarse-grained approximation of the full hydrodynamic treatment. In such an interpretation, the parameter $w_0$  incorporates in an effective way some average information about the hydrodynamic modes and could for instance be taken as proportional to the ratio of the permeability to the viscosity $\kappa/\eta$, with an explicit dependence on the pore width $H$ (see above). However, in what follows, we neglect such remains of the hydrodynamic treatment and consider $w_0$ as independent of $H$. (As a result, we do not expect to recover the proper trend for the front velocity as a function of the pore width and shall not study this aspect of the problem.)

As discussed in Refs.\cite{GPDM2003,M2008}, Eq. (\ref{evol}) may be recast into the discrete version of a Cahn-Hilliard\cite{CH1958} or dynamic density functional theory\cite{MT1999,AE2004} equation. The characteristic feature of the DMFT is that the mobility coefficient has an explicit expression and is related to the local site densities via the Metropolis transition probabilities. It is also worth noting that a coarse-grained continuity equation with a current $J$ proportional to the gradient of a chemical potential ({\it i.e.} a generalized Cahn-Hilliard equation with a Ginzburg- Landau type free energy) is at the core of the phase-field models that have been developed recently for studying spontaneous and forced-flow imbibition in disordered porous solids\cite{DREAMA2000,ADR2004,PH2006}. In such a continuum (mesoscopic) approach, the dry and wetted states are considered as two ``phases" of the system ($\phi=\pm 1$ where $\phi$ is the locally conserved field) and the influence of disorder on the capillary force and the viscous drag are included in a locally random, but concentration-independent, mobility. Hydrodynamic effects are thus hidden in this effective quantity.

An important feature of the DMFT, which will allow us to make a connection between the properties of the solution of the evolution equation and the adsorption isotherms in the same system, is that the site densities obtained from the minimization of the {\it static} Helmholtz free energy ${\cal F}[\{\rho_i\}]$ in the canonical ensemble (or from the minimization of the grand-potential $ \Omega={\cal F}-\mu\sum_{i} \rho_i $ in the grand canonical ensemble) are steady-state  solutions  of Eq. (\ref{evol}). Indeed, ${\cal F}$ in the mean-field approximation is given by
\begin{equation}
\label{Freeener}
\beta {\cal F}[\{\rho_i\}]=\sum_{i} [\rho_i \ln \rho_i + (1-\rho_i)\ln (1-\rho_i)] -\beta w_{ff}  \sum_{<ij>} \rho_i \rho_j  -\beta w_{sf}  \sum_{i,\: surface} \rho_i  \, ,
\end{equation}
and the equilibrium (or at least metastable) $\rho_i$'s  that satisfy the set of coupled equations
\begin{equation}
\frac{\rho_i}{1-\rho_i}=\exp [\beta (\mu-E_i)] 
\end{equation}
are also solutions of Eq. (\ref{evol}) when the left-hand side is set to zero\cite{M2008}.

In practice, we have integrated Eq. (\ref{evol}) via Euler's method ({\it i.e.} $\rho_i(t+\Delta t)=\rho_i(t)-\Delta t\sum_{j/i}J_{ij}(t)$) by using the dimensionless time step $w_0 \Delta t=0.2$. This value was small enough to ensure the stability of the solution in all  cases and it actually corresponds to a very small variation of the site densities (which leads to a serious technical problem and makes the numerical calculations rather time-consuming). For the  presentation of the results, we have used a larger time scale, $t_0=10^5 w_0 \Delta t$.

\section{Results and discussion}

Most of the results presented in this section correspond to the case of a ``dry" capillary, where the liquid in the reservoir is initially  in equilibrium with the vapor in the pore at $\mu=\mu_{sat}$.  This is realized by fixing the density of the fluid at the saturated liquid density $\rho_l(T)$ in the first (left) layer in the $z$ direction and at the saturated vapor density $\rho_g(T)$ in the rest of the pore. Then, at time $t=0$, we let the system evolve according to Eq. (\ref{evol}) and the liquid enters the pore. This is the case considered in the recent MD simulations\cite{DMB2007,CBD2008,AK2009} where the capillary walls are initially taken lyophobic and then switched to lyophilic at $t=0$.  As in these simulation studies, the pore in our set-up has its right end closed by a wall that prevents the fluid from escaping. An alternative set-up consists in starting from a  ``gas" configuration inside the pore with the adsorbed films along the walls already formed. 
We shall also briefly consider this second situation as it is as well encountered in experiments with nanocapillaries\cite{HGSKV2007,N2006}. 

The calculations are done at the reduced temperature $T^*=k_BT/w_{ff}=1$  (the bulk critical temperature is $T^*_c=1.5$)  for which $\rho_g\approx 0.07$ and $\rho_l\approx 0.93$. Unless explicitly stated, the pore width $H$ is fixed at $30$ lattice constants ({\it i.e.} $28$ fluid layers). For comparison, some calculations for other values of $H$ have also been performed. 

\begin{figure}[hbt]
\begin{center}
\includegraphics[width=8cm]{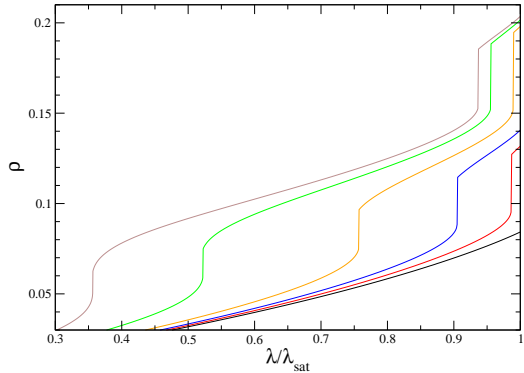}
\caption{\label{Fig2} (Color on line) Grand-canonical adsorption isotherms for $T^*=1$ in a pore of width $H=30$ for various values of the wettability parameter $\alpha$. From right to left: $\alpha=0.8, 0.9, 1, 1.2, 1.6, 2$ (for $\alpha=0.8$, the part of the isotherm  that extends beyond bulk saturation is not shown). The figure focuses on the range  of $\lambda/\lambda_{sat}$ associated to the formation of two-dimensional layers. The jump in density associated with capillary condensation is not shown.}
\end{center}
\end{figure}

For future reference, it is instructive to first examine the influence of the wettability parameter $\alpha$ on the  adsorption isotherms obtained from the free-energy, Eq. (\ref{Freeener}), in the grand-canonical ensemble. These are shown in Fig. \ref{Fig2} as a function of the relative activity $\lambda/\lambda_{sat}=\exp[\beta(\mu-\mu_{sat})]$. As usual, these isotherms have been obtained by starting from a low relative activity and increasing $\lambda$  in small steps.
This type of step-like isotherms is well documented in the literature (see {\it e.g.} \cite{BME1987}) and our primary interest here is in the two-dimensional layers of different thicknesses that form successively as $\alpha$ increases. Note that some of them are actually metastable since the capillary condensation (not shown in the figure) occurs before the layering transition (for instance, this is the case of the second layering transition for $\alpha =1.2$). This distinction, however, is irrelevant for the dynamic evolution during spontaneous imbibition as will be seen below.

As a necessary ingredient to the classical description of imbibition, it is important to know the variation of the static contact angle $\theta$ with $\alpha$. The contact angle is computed from Young's equation using the interfacial tensions on an infinite planar surface obtained from the mean-field free energy in Eq. (\ref{Freeener}). We refer the reader to Ref.\cite{M2008b} for more details on the calculation and to Fig. 2 in Ref.\cite{EM2010} for the results at $T^*=1$.  (Note that $\cos \theta=0$ for $\alpha=0.5$ due to the symmetry of the nearest-neighbor lattice-gas Hamiltonian\cite{PSW1982,KMRST2001}; note also that the weak dependence of $\theta$ on the direction of the interface\cite{SBB2005} is neglected in this calculation.) At this temperature, the wetting transition as a function of $\alpha$ is first-order and occurs slightly below $\alpha=0.9$. If one neglects the effect of confinement, this explains the behavior of the $\alpha=0.8$ isotherm in Fig. \ref{Fig2}, with no layering transition before $\lambda=\lambda_{sat}$.

\subsection{Dry capillary}

\begin{figure}[hbt]
\begin{center}
\includegraphics[width=7.5cm]{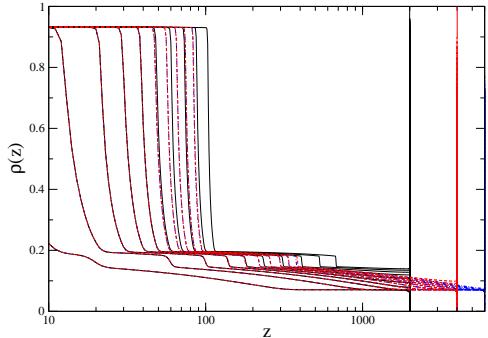}
\caption{(Color on line) \label{Fig3} Profiles of the average fluid density $\rho(z)$ in a pore of width $H=30$  as a function of time for $\alpha= 2$ and three different system lengths $L=2000$ (black solid lines), $L=4000$ (red dashed lines) and $L=6000$ (blue dot-dashed lines). From left to right, the profiles correspond to $t/t_0=4^2, 8^2,...20^2$. The distance $z$ is put on a logarithmic scale, so that $z=4000$ and $z=6000$ are virtually indistinguishable on the scale of the figure.}
\end{center}
\end{figure}

We first discuss an important methodological issue that must be considered in numerical studies of an initially dry capillary. It  concerns the length $L$ of the system. As usual, $L$ must be chosen large enough to avoid undesirable finite size or edge effects. When precursor films are present, it turns out that this is a much more severe requirement than for standard calculations of adsorption isotherms. This is illustrated in Fig. \ref{Fig3} where we compare the time evolution of the average fluid density profile $\rho(z)$ for $\alpha=2$ and three different pore lengths, $L=2000,4000$ and $6000$. 
The characteristic features of the profile will be discussed in more detail below but one can see at once that the advance of the main interface is modified as soon as the average fluid density at $z=L$ becomes significant (typically $\rho(L)\gtrsim 0.1$) due to the fact that the precursor film has already reached the pore end: this results in a significant increase in the velocity of the interface. The figure shows that for $L$  as large as $4000$ lattice constants (which is eventually the length used in our calculations), the numerical results are biased if the imbibition time exceeds $t\approx 400\: t_0$. Note that the main interface at this time has only reached a distance $z\approx 86$, much smaller than $L$ ! The length of the capillary is therefore an issue that must be treated very carefully in this type of calculation\cite{note1}.

\begin{figure}[hbt]
\begin{center}
\includegraphics[width=12cm]{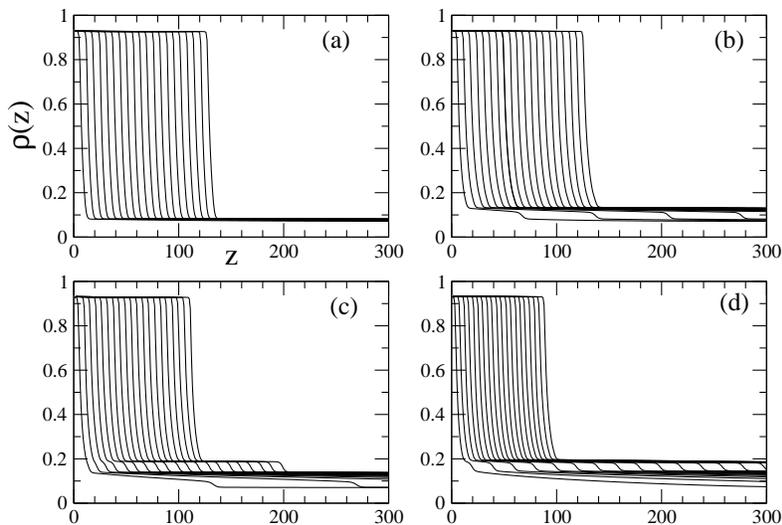}
\caption{\label{Fig4} Profiles of the average fluid density $\rho(z)$ in a pore of width $H=30$ ($L=4000$) as a function of time for different values of $\alpha$: (a) $0.8$, (b) $1$, (c) $1.2$, and  (d) $2$. The successive curves correspond to $t/t_0=1,2^2,3^2...20^2$.}
\end{center}
\end{figure}

We now discuss the main results of our calculations. Fig. \ref{Fig4} shows the time evolution of the average fluid density profile $\rho(z)$ in the capillary for different values of $\alpha$. 

In all cases, the observed equidistance between the successive profiles at times $t/t_0=1,2^2,3^2,4^2,...20^2$ shows that the main front, which corresponds to the largest variation in the average density and therefore can be associated to the liquid-gas interface in the center of the pore, advances like $\sqrt{t}$ (see also Fig. 5). In the complete wetting regime ($\alpha=1,1.2$ and $2$), the main front is preceded by a smaller front that advances faster, but also with a $\sqrt{t}$ law (for $\alpha=1.2$ there are actually two small fronts, the first one advancing faster than the second, see below). These fronts can be associated  to thin precursor films propagating along the pore walls.  For $\alpha=0.8$, which corresponds to a partial wetting situation, there is no precursor. On the whole, these observations are in line with the results of the recent off-lattice MD and lattice-Boltzmann simulations\cite{DMB2007,C2008}. The dependence on $\alpha$, however, is nontrivial and deserves a more detailed investigation. 

\begin{figure}[hbt]
\begin{center}
\includegraphics[width=9cm]{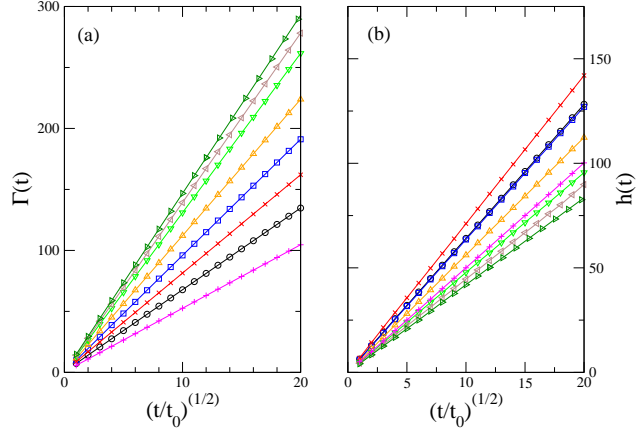}
\caption{\label{Fig5} (Color on line) (a) Total fluid uptake in the capillary as a function of the square root of imbibition time for various values of $\alpha$  (from bottom to top: $\alpha=0.7, 0.8, 0.9, 1, 1.2, 1.6, 2, 3$. The solid lines correspond to a $\sqrt{t}$ fit. (b) Advance of the liquid-gas meniscus. The symbols are the same as in (a) but the order of the curves is different (at the scale of the figure the results for $\alpha=0.8$ (circles) and $1$ (squares)  are almost indistinguishable). In both figures, the transient behavior for $t/t_0 <1$ is not shown.}
\end{center}
\end{figure}

\begin{figure}[hbt]
\begin{center}
\includegraphics[width=10cm]{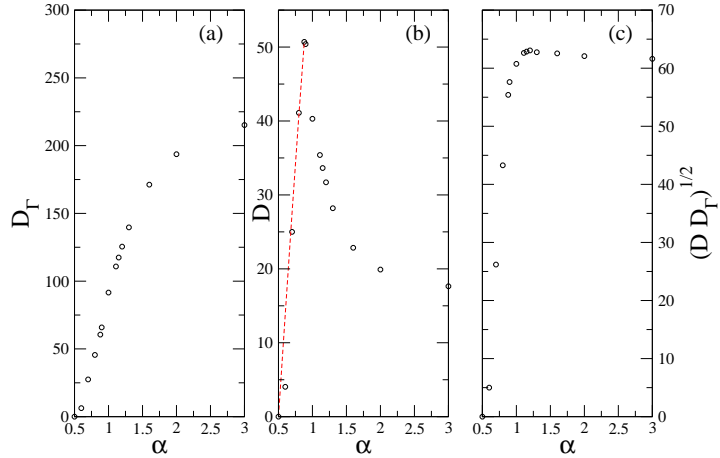}
\caption{ \label{Fig6} (Color on line)  Effective diffusion coefficients $D_{\Gamma}$ and $D$ for the time evolution of the total uptake (a) and the advance of the meniscus (b) as a function of the wettability parameter $\alpha$. In (b) the red dashed line corresponds to the formula $D=D(\alpha=0.88)\cos\theta$ (with $\theta(0.88)\approx 0$). In (c) the product  $(DD_{\Gamma})^{1/2}$ is shown to become approximately constant in the complete wetting regime, in agreement with Eq. (\ref{EqDDgamma}).}
\end{center}
\end{figure}

There is indeed an interesting difference between the evolution with time of the total fluid uptake  and that of the position of the meniscus. This is illustrated in Fig. \ref{Fig5} for values of $\alpha$ ranging from $0.7$ to $3$. The uptake (per unit lateral surface) is defined as $\Gamma(t)=\int [\rho(z,t) -\rho(z,t=0)]\:dz$, where we have subtracted the initial number of fluid molecules as we are only interested in the filling process for $t>0$. 
One can see in Figure \ref{Fig5}a that the uptake increases with $\alpha$ and appears to saturate for large values of $\alpha$ (well above the transition to complete wetting). The fit to a $\sqrt{t}$ law is excellent so that one can define an effective diffusion coefficient $D_{\Gamma}=\Gamma(t)^2/(t/t_0)$ whose variation with $\alpha$ is shown in Fig. \ref{Fig6}a.  On the other hand, the behavior displayed in Fig. \ref{Fig5}b for the advance of the meniscus appears at first sight surprising. Although the results can be again very well fitted to a $\sqrt{t}$ law, the variation of the velocity with $\alpha$ is not monotonic. The effective diffusion coefficient defined by $D=h(t)^2/(t/t_0)$ first increases (from $\alpha=0.7$ to $\approx 0.88$), then decreases, and finally saturates, as shown in Fig. \ref{Fig6}b. 

\begin{figure}[hbt]
\begin{center}
\includegraphics[width=7cm]{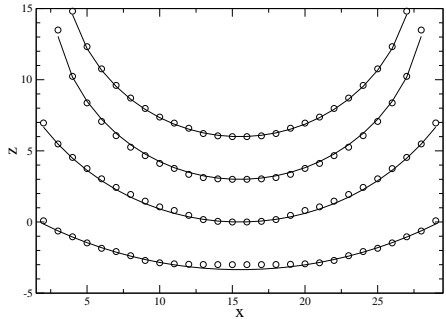}
\caption{  \label{Fig7} Meniscus profiles for different values of $\alpha$ at $t/t_0=250$. From bottom to top  $\alpha=0.7, 0.8, 1, 1.2$. For clarity, the curves are arbitrarily shifted vertically by a constant and the wetting layers adjacent to the walls (when present) are not shown. The solid lines correspond to a semi-circular fit. From this fit, we obtain $\cos \theta = 0.48, 0.83$ for $\alpha=0.7, 0.8$, to be compared with the values, $0.54, 0.805$, computed on a flat interface\cite{EM2010}; in both calculations, $\cos \theta \simeq 1$ for $\alpha \gtrsim 0.9$.}
\end{center}
\end{figure}

It is clear that the first regime in which the velocity of the meniscus increases with $\alpha$ coincides with the situation of partial wetting.  Moreover, for $0.7\lesssim \alpha\lesssim 0.88$, the effective diffusion coefficient is reasonably proportional to $\cos \theta$, in agreement with the macroscopic picture for the driving capillary force. For smaller values of $\alpha$, however, the relation between $D$ and $\cos \theta$ is no more linear, as if the actual driving force were smaller; since we only consider here the asymptotic steady state and no variation of the contact angle with velocity is observed (apart from small fluctuations\cite{note2}), the concept of dynamic contact angle cannot be invoked to explain the observed deviations\cite{note3}. They are more likely related to the deformation of the meniscus which can be seen in Fig. \ref{Fig7} where some typical interface profiles at the same imbibition time are shown. The interface is here defined as the density  isocontour $\rho(x,z)=0.5$ (the intrinsic width of the interface being equal to a few lattice constants, which is standard at this temperature). For $\alpha\gtrsim 0.7$, the menisci can be well fitted to a semi-circle with radius $R$  and the contact angle extracted from the tangent at the wall ({\it i.e.} $\cos \theta\approx H/2R$) is in reasonable agreement with the static angle computed on a flat interface\cite{EM2010}: see the caption of Fig. 7. On the other hand, for $\alpha\lesssim 0.7$, some flat portion appears in the center of the pore, a feature which may possibly be attributed to lattice artifacts that become more and more significant as $\theta$ approaches $\pi/2$.

The unexpected decrease of $D$ with $\alpha$ in the complete wetting regime is related to the presence of precursor films\cite{note4}.  Indeed, because of mass conservation, the velocities of the meniscus and the precursor front are not independent. A semi-quantitative argument is developed in the Appendix where we investigate the coarse-grained behavior of the DMFT evolution equation on a scale where the structure of the interface may be neglected. In particular, we derive the relation
\begin{equation}
\label{EqDDgamma}
\sqrt{D D_{\Gamma}}\propto \delta \mu
\end{equation}
where $\delta \mu=\mu_{sat}-\mu^*$ and $\mu^*$ is the value of the chemical potential at the main interface. $\delta \mu$ is a small quantity  that does not vary much in the complete wetting regime ($\delta \mu/w_{ff}\approx 0.023\pm 0.002$) and is approximately equal to $(\rho_l-\rho_g)\gamma_{lv}/R$, in agreement with the  macroscopic Gibbs-Thomson equation. As shown in  Fig. \ref{Fig6}c, the product $\sqrt{D D_{\Gamma}}$ is indeed constant in the whole range $\alpha=1-3$ with an error less than $5\%$ (for $\alpha=0.9$ the product is smaller, possibly because the complete wetting regime is not fully established). It is remarkable that the product of the two effective diffusion constants $D$ and $D_{\Gamma}$ is only controlled by $\delta \mu$. Since the speed of the {\it whole} imbibed fluid, including the precursor film and beyond, increases with $\alpha$ (which is expected on  physical ground), Eq. (\ref{EqDDgamma}) tells us that the speed of the meniscus must decrease.

\begin{figure}[hbt]
\begin{center}
\includegraphics[width=9cm]{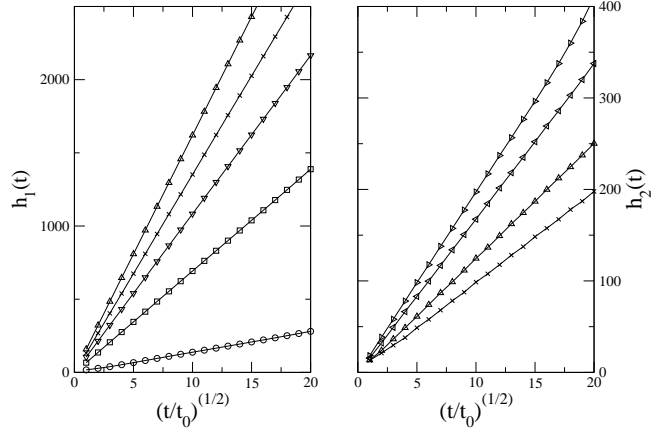}
\caption{  \label{Fig8} The positions $h_1(t)$ and $h_2(t)$ of the precursor films as a function of the square root of time for several values of $\alpha$ in the complete wetting regime. For $h_1(t)$: $\alpha=0.9, 1, 1.1, 1.2, 1.3$ from bottom to top; for $h_2$: $\alpha=1.2, 1.3, 1.6, 2$ from bottom to top. The solid lines correspond to a $\sqrt{t}$ fit.}
\end{center}
\end{figure}

We now focus on the dynamics of the precursor films. Figure \ref{Fig8} shows the position of the precursor films as a function of  $\sqrt{t/t_0}$ for increasing values of $\alpha$.  As was observed in Fig. \ref{Fig4} there may exist one or two moving fronts: $h_1(t)$ (resp. $h_2(t)$) denotes the position of the front corresponding to the filling of the first (resp. second layer) adjacent to the pore walls (the fronts are defined after averaging over the transverse direction $x$ and thus incorporate the layers adjacent to the two walls and the gas in between). The first front  is not well defined for $\alpha \gtrsim 1.3$.
One can see that the velocity of the fronts increases with $\alpha$  and that the first one develops a faster dynamics when both fronts coexist.  (The motion is much faster than that of the meniscus, which is why we were forced to consider very large systems.) In all cases,  the fit to a $\sqrt{t}$ law is excellent, in agreement with the coarse-grained analysis performed in the Appendix (see Eq. (\ref{A16})).

\begin{figure}[hbt]
\begin{center}
\includegraphics[width=9cm]{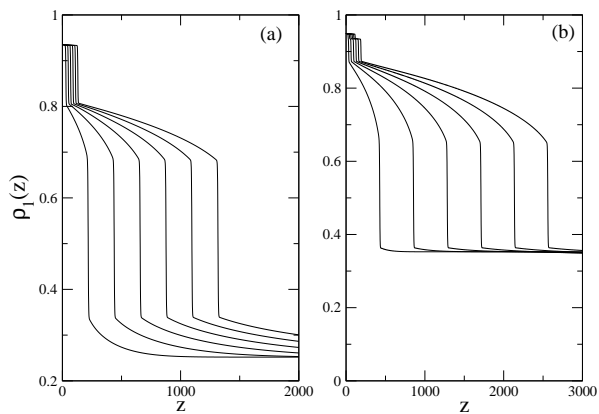}
\caption{ \label{Fig9} Density $\rho_1(z)$ in the first layer for $\alpha=1$ (a) and $\alpha=1.2$ (b) at times $t/t_0=10, 40, 90, 160, 250, 360$. In (a) the large and small jumps are associated to the passage of the precursor front and the passage of the meniscus, respectively. In (b) there is an intermediate jump associated to the passage of the second front displayed in Fig. \ref{Fig4}c.}
\end{center}
\end{figure}

It is important to note that the precursor films do not advance on completely dry walls. Indeed, as soon as one lets the system evolve, there is a very quick redistribution of the gas inside the pore due to the attraction exerted by the walls. This transient process has no relation with the progression of the liquid from the reservoir and it thus establishes a new inhomogeneous ``canonical" equilibrium (accordingly, the chemical potential at $z=L$  becomes lower than $\mu_{sat}$ as shown in Fig. \ref{Fig13} in the Appendix). This effect is not visible in Fig. \ref{Fig4} which only displays the average fluid density $\rho(z)$, but it can be seen in Fig. \ref{Fig9} where we plot the time evolution of the density $\rho_1(z)$ in the first layer for $\alpha=1$ and $1.2$.
For $t/t_0=10$, one finds $\rho_1(1000)\approx 0.25$ for $\alpha=1$ and $\approx 0.36$  for $\alpha=1.2$, which are both values significantly larger than the density $\rho_g\approx 0.07$  that was imposed at the beginning of the calculation. As expected, this initial density increases with $\alpha$ and we have checked that it also increases with the pore width as there are more particles than can be ``pumped"  from the center of the pore to be adsorbed on the walls. This inhomogeneous ``gas" configuration should thus be considered as the actual initial state of the pore before imbibition begins. 

\begin{figure}[hbt]
\begin{center}
\includegraphics[width=8cm]{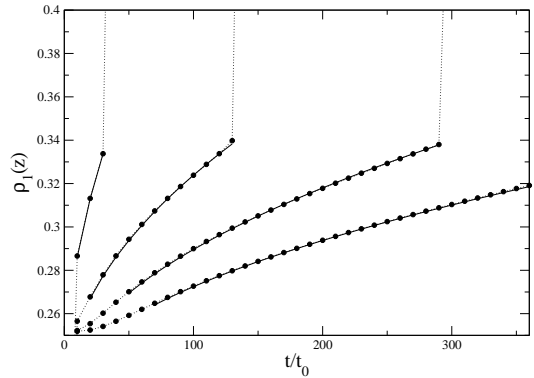}
\caption{  \label{Fig10} Time evolution of the density in the first layer at a fixed distance from the pore entrance (from left to right, $z=400,800, 1200, 1600$)  for $\alpha=1$. The solid lines correspond to the fit $\rho_1(z,t)\approx 0.221+8.23\sqrt{(t/t_0)/z^2}$.}
\end{center}
\end{figure}

Fig. \ref{Fig9} shows that the filling dynamics of the first layer can be divided into distinct stages. For instance, for $\alpha=1$, there is first a gradual filling, then a  large jump corresponding to the passage of the precursor front,  a new gradual filling, and  finally a small jump corresponding to the passage of the meniscus. (For $\alpha=1.2$  the first stage is absent but there is an additional intermediate jump that corresponds to the passage of the second front ahead of the meniscus seen in Fig. \ref{Fig4}c .) The first stage of this process is analyzed in Fig. \ref{Fig10} which shows the time evolution of $\rho_1(z)$ at a {\it fixed} distance $z$ for $\alpha=1$. The gradual filling of the layer is well fitted by the formula $\rho_1(z,t)=a+b\sqrt{t/z^2}$ so that this process is also diffusive. 

Remarkably, the values of the average density $\rho$ in the pore before and after the passage of the precursor front are in good agreement with the values of the density observed before and after the corresponding layering transitions in the adsorption isotherms : for example, these densities in Fig. \ref{Fig4} are respectively $0.083$ and $0.110$ for $\alpha=1$, to be compared to $0.087$ and $0.115$ in Fig. \ref{Fig2}; for $\alpha=2$, the density jumps from $0.148$ and $0.180$ at the passage of the front to be compared to $0.152$ and $0.185$ before and after the second layering transition in Fig. \ref{Fig2}. The corresponding chemical potentials are also very close (for instance, $\mu_1/w_{ff}\approx -3.11$ in Fig. \ref{Fig13}b and $\mu_1/w_{ff}=-3+T^* \ln(\lambda_1/\lambda_{sat})\approx -3.10$ in Fig. \ref{Fig2} for $\alpha=1$). This illustrates the fact that the same states are encountered during imbibition and adsorption processes, the possible metastability of a state with respect to capillary condensation being irrelevant dynamically. 

One could thus interpret the jump associated to the passage of the precursor front as a {\it dynamic} layering transition that occurs when $\rho$ (or $\rho_1$) reaches a certain critical value (for example, $\rho_1\approx 0.34$ for $\alpha=1$).  In this interpretation, the speed of the precursor is directly related to the time needed to reach this critical density. Since the initial density in the layer increases with the pore width, as noted above, this time gets shorter and the precursor moves faster as $H$ increases,  which is indeed observed in our calculations. Moreover, the intermediate stage corresponding to the gradual filling of the layer may be totally absent. Knowing the initial configuration of the fluid in the pore is therefore important to understand the dynamical behavior of the film, which is itself related to the dynamical behavior of the meniscus, as shown by Eqs. (\ref{A17}) in the Appendix.  (Of course, the viscous drag also varies with the pore width, an effect which is not included, even in an effective manner, in the present treatment where we have taken $w_0$ as independent of $H$.)

\subsection{Prewetted capillary}

We now consider the situation of a prewetted capillary.  The pre-existing films coating the pore walls were created by gradually increasing the relative activity $\lambda/\lambda_{sat}$  from a low value up to $1$, and one or two monolayers were thus obtained depending on the value of $\alpha$, as shown in Fig. \ref{Fig2}.  Of course, these thin films are metastable since the true equilibrium situation at $\mu_{sat}$ corresponds to the pore entirely filled with liquid and, in any case, capillary condensation occurs before $\mu_{sat}$. This feature, however, is not important for the present purpose and this situation may possibly occur in actual experiments with nanoporous disordered solids, depending on the timescales involved in capillary condensation and imbibition\cite{N2006}.

\begin{figure}[hbt]
\begin{center}
\includegraphics[width=12cm]{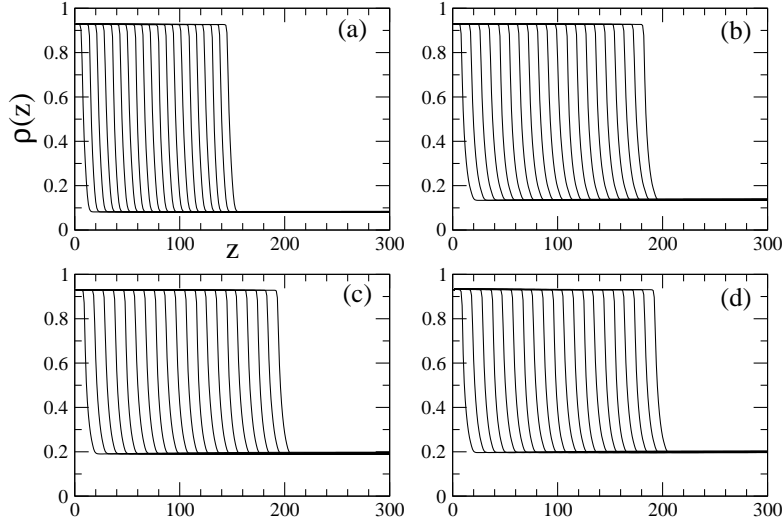}
\caption{ \label{Fig11} Same as Fig. \ref{Fig4} for the prewetted capillary.}
\end{center}
\end{figure}

\begin{figure}[hbt]
\begin{center}
\includegraphics[width=9cm]{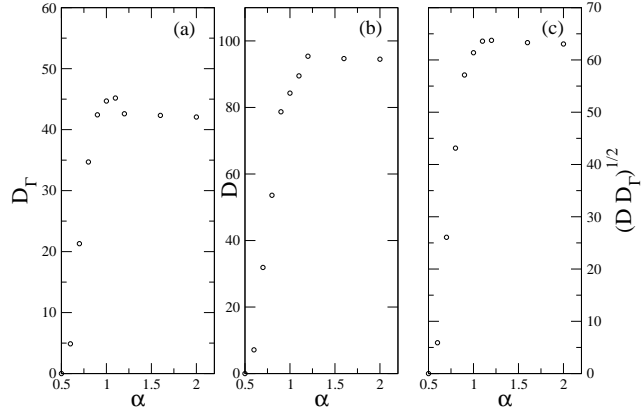}
\caption{ \label{Fig12}  Same as Fig. \ref{Fig6} for the prewetted capillary.}
\end{center}
\end{figure}

Figure \ref{Fig11} shows the time evolution of the average fluid density profile $\rho(z)$ for various values of $\alpha$. Comparing with Fig. \ref{Fig4}, one observes the absence of precursor fronts, as expected.  The advance of the interface again follows a $\sqrt{t}$ law, but the speed of the meniscus now increases with $\alpha$ like the total fluid uptake, as shown in Fig. \ref{Fig12}. Moreover, both quantities saturate more rapidly with $\alpha$ than in the presence of  precursor films. The whole dynamical behavior is therefore quite simpler. 

It is generally expected that the meniscus moves faster in a prewetted capillary as the frictional force is reduced\cite{note5}. Comparing Figs. \ref{Fig11} and \ref{Fig12} with  \ref{Fig5} and \ref{Fig6} shows that this behavior is also observed in our calculations, but this is now due to the fact that the meniscus no more competes with the precursor film for  the liquid coming from the reservoir. The diffusion coefficient for the total uptake on the other hand is smaller than in the previous set-up (note that the product $\sqrt{DD_{\Gamma}}$ is almost the same in the two cases because Eq. (\ref{EqDDgamma}) still holds and $\delta \mu$ does not really change). 

\section{Summary and conclusion}

To summarize, we have investigated spontaneous imbibition in a narrow slit pore in the framework of a dynamic lattice mean-field theory that explicitly includes liquid conservation but neglects momentum transport. The theory is able to reproduce the main characteristic feature of imbibition in the quasistatic regime, namely the constant slowing down of the interface described by the macroscopic $\sqrt{t}$ Lucas-Washburn law.  When the capillary is initially dry and the solid-fluid interaction is sufficiently attractive (so that the static contact angle is zero),  we have found that thin precursor films move ahead of the meniscus, following also a $\sqrt{t}$ law, as also reported by recent simulation studies\cite{DMB2007,C2008,CBD2008}. The jump in the average fluid density in the pore that is associated to the passage of the precursor front may be interpreted as a dynamic layering transition, whose existence can be predicted from the corresponding adsorption isotherms. Due to mass conservation, the meniscus and the precursor compete for the liquid coming from the reservoir, which induces a nontrivial dependence of the speed of the interface on the strength of the solid-fluid interaction. On the other hand, the total fluid uptake increases monotonously.
This contrasts with the simpler behavior observed in a capillary where thin films adsorbed on the pore walls are pre-existing: in this case, both the speed of the meniscus and the fluid uptake increase with the solid-fluid interaction and rapidly saturate in the complete wetting regime.

The extent to which the above predictions may be tested in actual experiments or in simulations remains an open question. We have emphasized that the system length is an issue that must be carefully  treated in simulations  when precursor films are present. At a different level, since our treatment has drastically simplified the complexity of the problem, the phenomena revealed by our calculations may be hidden or significantly altered. For instance, we have neglected a possible dependence of the molecular friction coefficient on the strength of the solid-fluid interaction in the first layers adjacent to the pore walls. This dependence has been invoked to explain the nonmonotonic variation of the film spreading rate on solid substrates with different surface energies\cite{VVOCC1998}. This could be included in the present theory by modifying the jump rate parameter $w_0$ in the first layer (for instance, by taking  $w_1=w_0 \exp(-C\alpha)$ where $C$ is some constant). By reducing the speed of the precursor for large $\alpha$, this would in turn accelerate the speed of the meniscus.
The absence of hydrodynamics is of course another very serious limitation if one wishes to go beyond a mere account through an effective elementary time scale. 
In principle, momentum transport could be incorporated at a coarse-grained level by coupling the density to a velocity field satisfying the appropriate Navier-Stokes equation (see {\it e.g.} \cite{JV1996,BJ2008}). This, however, would considerably complicate the numerical treatment and make  the study of more complex geometries encountered in disordered porous media virtually impossible.

\acknowledgments
This work was supported by ANR-06-BLAN-0098.

\appendix 

\section{Coarse-grained description  of the DMFT evolution equation for imbibition}

We start with the DMFT evolution equation, Eq. (\ref{evol}), namely,
\begin{equation}
\label{A1}
\frac{\partial\rho_i}{\partial t}=-\sum_{j/i}\left[w_{ij}\rho_i(1-\rho_j)-w_{ji}\rho_j(1-\rho_i)\right] \ .
\end{equation}

Due to the symmetry, the $y$ direction is irrelevant in the problem which is effectively two-dimensional in the $(x,z)$ plane.
We consider the sum of the $\rho_i$'s over the $x$ direction at $z$ constant. It is then easy to show that for all nearest neighbors of $i$ in the $z=$ constant plane,
\begin{equation}
\label{A2}
\sum_{<ij>,\:z \:\mbox{constant}}[w_{ij}\rho_i(1-\rho_j)-w_{ji}\rho_j(1-\rho_i)]=0 \ .
\end{equation}
Then
\begin{align}
\label{A3}
\frac{\partial }{\partial t} \sum_x \rho(x,z)&=-\sum_x\Big\{[w_x(z,z+1)(1-\rho(x,z+1))+w_x(z,z-1)(1-\rho(x,z-1))]\rho(x,z)\nonumber\\
&-[w_x(z+1,z)\rho(x,z+1)+w_x(z-1,z)\rho(x,z-1)](1-\rho(x,z))\Big\}
\end{align}
where we have defined $w_x(z,z')=w_{ij}$ with $i = (x,z)$ and $j = (x,z')$. 

On physical ground, we expect that the energy $E(x,z)$ increases and the density $\rho(x,z)$ decreases with increasing $z$ at constant $x$. Considering the definition of the rates, it is then found that
\begin{equation}
\label{A4}
\left\{
\begin{array}{ll}
w_x(z,z-1)=w_0,       & w_x(z,z+1)=w_0e^{-\beta[E(x,z+1)-E(x,z)]}\\
w_x(z+1,z)=w_0,      & w_x(z-1,z)=w_0e^{-\beta[E(x,z)-E(x,z-1)]} \ .
\end{array}\right.
\end{equation}
\begin{figure}[hbt]
\begin{center}
\includegraphics[width=10cm]{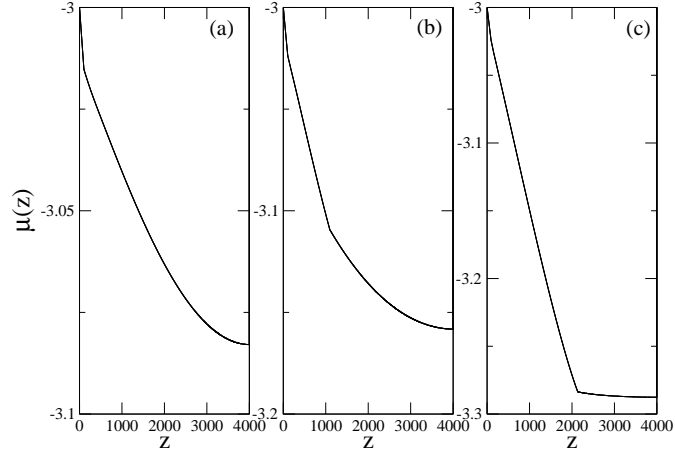}
\caption{  \label{Fig13} Chemical potential profile in the center of the pore (solid line) and in the first two layers adjacent to the pore walls (dashed and dotted-dashed lines) for $t/t_0=250$ and $\alpha=0.8$ (a), $1$ (b), and $1.2$ (c). On the scale of the figure the three curves are indistinguishable.}
\end{center}
\end{figure}
We define the local chemical potential $\mu(x,z)$ as an extension of Eq. (16):
\begin{equation}
\label{A5}
\beta \mu(x,z)=\ln\frac{\rho(x,z)}{1-\rho(x,z)}+\beta E(x,z) \ .
\end{equation}
Then,
\begin{align}
\label{A6}
\frac{\partial}{\partial t} \sum_x \rho(x,z)=&-w_0\sum_x\Big\{(e^{-\beta[\mu(x,z+1)-\mu(x,z)]}-1)\rho(x,z+1)(1-\rho(x,z))\nonumber\\
&-(e^{-\beta[\mu(x,z)-\mu(x,z-1)]}-1)\rho(x,z)(1-\rho(x,z-1))\Big\} \ .
\end{align}
Except possibly in the vicinity of the interfaces, the chemical potential is expected to change smoothly and slowly. To take the continuum limit which corresponds to a phase-field model, one has to reinstall the lattice spacing ``a" and take the appropriate limit $a \rightarrow 0$. In any case, $\beta[\mu(x,z+a)-\mu(x,z)] \sim a \ll1$, which leads to
\begin{align}
\label{A7}
\frac{\partial }{\partial t} \sum_x \rho(x,z)&=\beta w_0\sum_x\Big\{[\mu(x,z+a)-\mu(x,z)]\rho(x,z+a)[1-\rho(x,z)]\nonumber\\
&\:\:\:\:\:\:\:\:\:\:\:\:\:\:\:\:\:\:\:-[\mu(x,z)-\mu(x,z-a)]\rho(x,z)[1-\rho(x,z-a)]\Big\} \nonumber\\
&=\beta w_0\sum_x\Big\{ [\mu(x,z+a)+\mu(x,z-a)-2\mu(x,z)]\rho(x,z)[1-\rho(x,z)]\nonumber\\
&\:\:\:\:\:\:\:\:\:\:\:\:\:\:\:\:\:\:\:+[\mu(x,z+a)-\mu(x,z)](1-\rho(x,z))[\rho(x,z+a)-\rho(x,z)]\nonumber\\
&\:\:\:\:\:\:\:\:\:\:\:\:\:\:\:\:\:\:\:-[\mu(x,z)-\mu(x,z-a)]\rho(x,z)[\rho(x,z)-\rho(x,z-a)]\Big\} \ .
\end{align}

In the continuum limit, $a\rightarrow 0$ and $w_0a^2\rightarrow {\tilde w}_0$, one finds
\begin{align}
\label{A8}
\frac{\partial }{\partial t} \sum_x \rho(x,z)&=\beta {\tilde w}_0\sum_x\Big\{\partial^2_z\mu(x,z)\rho(x,z)[1-\rho(x,z)]+\partial_z\mu(x,z)\partial_z\rho(x,z)[1-2\rho(x,z)]\Big\} \nonumber\\
&=\beta {\tilde w}_0 \sum_x\Big\{\partial^2_z\mu(x,z)\rho(x,z)(1-\rho(x,z))+\partial_z\mu(x,z)\partial_z\big[\rho(x,z)[1-\rho(x,z)]\big]\Big\} \nonumber\\
&=\beta {\tilde w}_0 \:\partial_z\Big\{\sum_x[\partial_z\mu(x,z)]\rho(x,z)[1-\rho(x,z)]\Big\}
\end{align}
When assuming  that $\mu(x,z)$ is independent of $x$, the evolution equation for the average density profile $\rho(z)=(1/H)\sum_x\rho(x,z)$ can finally be written as 
\begin{equation}
\label{A9}
\frac{\partial }{\partial t} \rho(z)=\partial_z\Big(M(z)\partial_z\mu(z)\Big)
\end{equation}
where $M(z)=\beta{\tilde w}_0(1/H)\sum_x \rho(x,z)[1-\rho(x,z)]$ is a local, density-dependent mobility (the time dependence of $\rho(z)$ and $\mu(z)$ is left here implicit). A phase-field description is thus recovered at a coarse-grained level because the variation of the local chemical potential must be smooth enough to allow the expansion of the exponential and more generally, the variation of all $z$-dependent quantities must be smooth enough to allow the expansion in the lattice spacing $a$.
Note that here $\rho(z)$ is a true fluid density profile and not simply a mathematical object identifying the phases as in usual phase-field models.

We now consider the physical situation encountered in the present study, in which the rapid variations of $\rho$ and $\mu$ are localized at $h(t)$ (the position of the main front) and, when precursor films are present, at $h_1(t)>h(t)$. For simplicity, we assume that there is only one precursor front as found for $\alpha =1$ and $\alpha=2$; the calculation can be easily generalized to the situation where two or more fronts are present. We use the above phase-field description, neglecting the details of the fluid near the interfaces that are assumed to be sharp, and we focus on  the region behind $h_1(t)$. For $h(t)\le z\le h_1(t)$ the averaged density $\rho(z)=(1/H)\sum_x\rho(x,z)$  thus accounts for the two films on the pore walls and the gas in between. Inspection of the numerical solution shows that (i) $\mu(x,z)$ is indeed almost independent of $x$ far from the interfaces, as can be see in Fig. \ref{Fig13}, (ii)
the mobility $M(z)$ is approximately piecewise constant in this region, $M(z)\approx M_l= \beta{\tilde w}_0\rho_l\rho_g$ for $0\le z<h(t)$ and $M(z)\approx M^*>M_l$ for $h(t)\le z<h_1(t)$.  Reinstalling explicitly the time dependence in Eq. (\ref{A9}), we thus have
\begin{equation}
\label{A10}
\frac{\partial }{\partial t} \rho(z,t)=\left\{
\begin{array}{ll}
M_l\:\frac{\partial^2}{\partial z^2} \mu(z,t) \ ,    & 0\le z<h(t)\\
\\
M^*\:\frac{\partial^2}{\partial z^2} \mu(z,t)\ ,      &h(t)\le z<h_1(t) \ .
\end{array}\right.
\end{equation}
At zeroth-order, a solution of this equation  is provided by a piecewise description with $\mu(z,t)$ linear in $z$ and $\rho(z,t)$ constant,
\begin{equation}
\label{A11}
\begin{array}{ll}
\mu(z,t)=\mu_{sat}-(\mu_{sat}-\mu^*)\frac{z}{h(t)} \\
\rho(z,t)=\rho_l 
\end{array} \Big\} \: \: \mbox{for $0\le z<h(t)$}
\end{equation}
and
\begin{equation}
\label{A12}
\begin{array}{ll}
\mu(z,t)=\mu^*-(\mu^*-\mu_1)\frac{z-h(t)}{h_1(t)-h(t)} \\
\rho(z,t)=\rho^{*}
\end{array} \Big\} \: \: \mbox{for $h(t)\le z<h_1(t)$}
\end{equation}
where $\rho^*$, $\mu^*$ and $\mu_1$ are considered as empirically fitted parameters (they can actually be evaluated by considering the adsorption isotherms, see Fig. 2). Note that this description is of course approximate and that corrections are present. For instance, as seen by blowing up the density profiles in Fig. 4, one finds that the density very slightly decreases between $0$ and $h(t)$ (as an immediate consequence of the decrease in $\mu$) as well as between $h(t)$ and $h_1(t)$. There is also a linear variation of the mobility $M(z)$. These corrections however are small and well within the uncertainty of the present coarse-grained picture. The situation is more serious in the region ahead of the most advanced front ($h_1(t)$ if there is a precursor film, $h(t)$ otherwise). Due to the attraction exerted by the pore walls, the chemical potential and the density are not simply those of the homogeneous gas and, as illustrated in Fig.  \ref{Fig13}, the chemical potential significantly varies with $z$ and strongly deviates from a linear behavior (as is also observed for the density and the mobility). Computing $\mu(z)$ and $\rho(z)$ in this region is out of the scope of the coarse-grained description but one can reasonably assume (and check numerically) that these quantities scale like $h(t)$ or $h_1(t)$, for instance $\mu(z)={\tilde \mu}(z/h_1(t))$  when there is a precursor film.

To determine the time evolution of the density profile for $0\le z\le h_1(t)$ we then integrate Eq. (\ref{A10}) over $z$ in infinitesimal intervals around $h(t)$ and $h_1(t)$. This yields
\begin{align}
\label{A13}
\int_{h(t)-\epsilon}^{h(t)+\epsilon} \frac{\partial \rho(z,t)}{\partial t} dz &=M^*\frac{\partial \mu(z,t)}{\partial z }\vert_{h(t)+\epsilon}-M_l\frac{\partial \mu(z,t)}{\partial z }\vert_{h(t)-\epsilon}\nonumber\\
&=M_l\frac{\mu_{sat}-\mu^*}{h(t)}-M^*\frac{\mu^*-\mu_1}{h_1(t)-h(t)}  \ ,
\end{align}
and thus, using the Heaviside form of the density profile at $z=h(t)$, 
\begin{equation}
\label{A14}
(\rho_l-\rho^*){\dot h}(t)=M_l\frac{\mu_{sat}-\mu^*}{h(t)}-M^*\frac{\mu^*-\mu_1}{h_1(t)-h(t)}  \ .
\end{equation} 
Similarly, integrating between $h_1(t)-\epsilon$ and $h_1(t)+\epsilon$:
\begin{equation}
\label{A15}
(\rho^*-\rho^{**}){\dot h}_1(t)=M^*\frac{\mu^*-\mu_{1}}{h_1(t)-h(t)}+M^{**}\frac{{\tilde \mu}'(1)}{h_1(t)} \ ,
\end{equation}
where $\rho^{**}$, $M^{**}$, and ${\tilde \mu}'(1)$ are  the average density, the mobility, and the derivative of ${\tilde \mu}(z/h_1)$ at $z=h_1(t)+\epsilon$, respectively. The solution of these two coupled differential equations is easily found  to be of the diffusive form,
\begin{align}
\label{A16}
h(t)&=\sqrt{Dt}\nonumber\\
h_1(t)&=\sqrt{D_1t} \ ,
\end{align}
with $D$ and $D_1$ solutions of the equations
\begin{align}
\label{A17}
(\rho_l-\rho^*)\sqrt{D}&=2M_l\frac{\mu_{sat}-\mu^*}{\sqrt{D}}-2M^*\frac{\mu^*-\mu_1}{\sqrt{D_1}-\sqrt{D}} \nonumber\\
(\rho^*-\rho^{**})\sqrt{D_1}&=2M^*\frac{\mu^*-\mu_{1}}{\sqrt{D_1}-\sqrt{D}}+2M^{**}\frac{{\tilde \mu}'(1)}{\sqrt{D_1}} \ ,
\end{align}
where $D_1>D$. Once the various parameters are known, $D$ and $D_1$ are easily computed. 

Finally, the time evolution of the total fluid uptake $\Gamma(t)=\int_0^L dz \:[\rho(z)-\rho(z,t=0)]$ is obtained from
\begin{align}
\label{A18}
{\dot \Gamma}(t)=\int_0^L dz \: \frac{\partial \rho(z,t)}{\partial t}=-[M(z) \frac{\partial \mu(z)}{\partial z}]_{z=0}=M_l\frac{\mu_{sat}-\mu^*}{h(t)} \ ,
\end{align}
where we have used the fact that the derivative of $\mu(z)$ vanishes at the right end of the pore when $L$ is large enough. This leads to
\begin{align}
\label{A19}
\Gamma(t)=\sqrt{D_{\Gamma}t}
\end{align}
with 
\begin{align}
\label{A20}
\sqrt{D_{\Gamma}}=2M_l\frac{\mu_{sat}-\mu^*}{\sqrt{D}} \ .
\end{align}

Eqs. (\ref{A16}), (\ref{A19}) and (\ref{A17}), (\ref{A20}) provide the full solution of the (simplified) problem.


\begin{thebibliography}{10}
\bibitem{note1} In particular, when looking at Figure 2 in Ref. \cite{DMB2007}, it is not certain that the capillary tube is long enough.
\end{thebibliography}
\end{document}